\documentclass[]{pasj01}
\usepackage{makeidx}

\makeatletter
\newcommand\@idxitem{\par\hangindent 40\p@}

\makeatother

\makeindex

\newcommand{\kms}{km~s$^{-1}$}

\newcounter{adc}

\begin{document}
	\Received{}
	\Accepted{}
	
	\title{Sunspot Structure\thanks{Can be cited as: Tiwari, Sanjiv K. and Hinode Review Team, ``Achievements of Hinode in the First Ten Years," 2018, PASJ, under review. \#Now with the affiliations 3 (LMSAL) and 4 (BAERI).}}
	
	\author{Sanjiv~K.~\textsc{Tiwari}\altaffilmark{1,2,3,4\#}}
	
	\altaffiltext{1}{NASA Marshall Space Flight Center, Mail Code ST 13, Huntsville, AL 35812, USA}
	\altaffiltext{2}{Center for Space Plasma and Aeronomic Research, University of Alabama in Huntsville, AL 35899, USA}
	\altaffiltext{3}{Lockheed Martin Solar and Astrophysics Laboratory, 3251 Hanover Street, Building 252, Palo Alto, CA 94304, USA}
	\altaffiltext{4}{Bay Area Environmental Research Institute, NASA Research Park, Moffett Field, CA 94035, USA}
	
	\email{tiwari@lmsal.com}
	
	
	\maketitle
	
	\begin{abstract}
	Sunspots contain multiple small-scale structures in the umbra and in the penumbra. Despite extensive research on this subject in pre-Hinode era multiple questions concerning fine-scale structures of sunspots, their formation, evolution and decay remained open. Several of those questions were proposed to be pursued by Hinode (SOT). Here we review some of the achievements on understanding sunspot structure by Hinode in its first 10 years of successful operation. After giving a brief summary and updates on the most recent understanding of sunspot structures, and describing contributions of Hinode to that, we also discuss future directions. This is a section (\#7.1) of a long review article on the achievements of Hinode in the first 10 years.
	\end{abstract}

\section{Introduction}
\label{sec71}
\begin{figure*}
	\includegraphics[trim=0cm 0cm 0cm 0cm, clip=true, width=\textwidth]{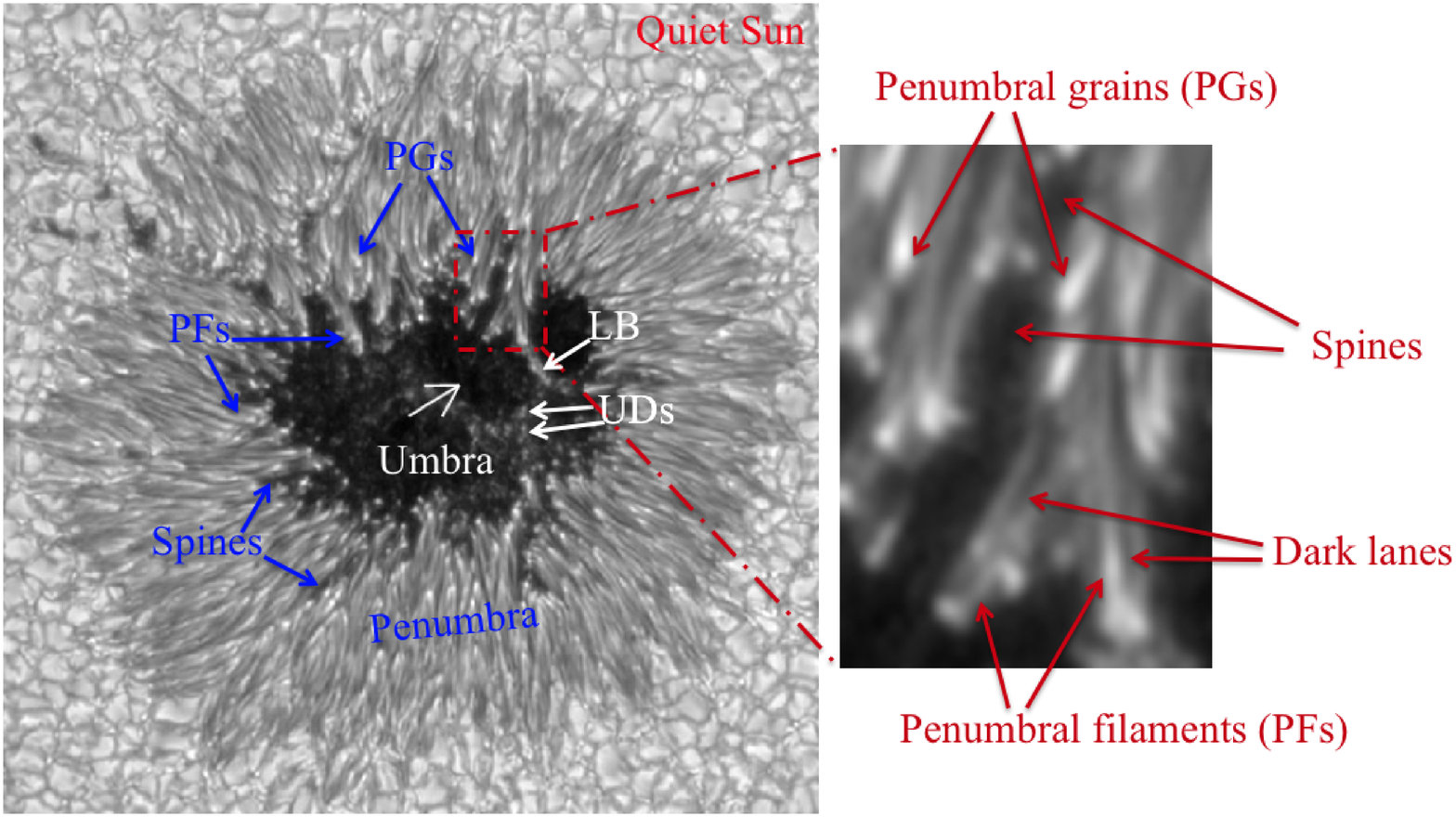}
	\caption{%
		Left: Continuum intensity image of a sunspot observed by Hinode/SOT-SP
		[reproduced from \citet{2015A&A...583A.119T} by permission of ESO]. Locations of a couple of umbral dots (UDs), penumbral filaments (PFs), spines, penumbral grains (which are actually heads of filaments) and a light bridge (LB), are pointed to by arrows. A larger arrow in the center of the sunspot umbra points to the direction of the solar disk center. The scale of the picture is $64'' \times 64''$. To clearly visualize the penumbral features (including dark lanes on penumbral filaments) a zoomed in view of a small FOV of the sunspot penumbra, outlined by dash-dotted box, is displayed in the right.
	}
	\label{7.1-fig1}
\end{figure*}

Sunspots, dark features on the surface of the Sun due to the suppressed convection owing to the presence of strong magnetic field in them, contain multiple small-scale structures in the central darkest part, the umbra, and in the less-dark region surrounding the umbra, the penumbra (figure \ref{7.1-fig1}). The magnetic, thermal, and flow structures of sunspots were extensively studied in the pre-Hinode era. But multiple questions pertaining to sunspot fine structure, their formation, evolution and decay, remained open, requiring a closer look. Some of these questions were proposed to be pursued by Hinode (SOT). For example, what are the internal structures of basic umbral and penumbral features (i.e., umbral dots, umbral dark area, light bridges, penumbral filaments, spines, penumbral bright grains) of sunspots and how are these basic umbral and penumbral structures formed and maintained? What drives the Evershed flow in sunspot penumbra in the photosphere and the inverse Evershed flow in sunspot penumbra in the chromosphere? How do the basic sunspot structures disintegrate in magnetic fragments and diffuse to the quiet Sun? How do moving magnetic features form and what is their role in sunspot decay? Are umbral dots, light bridges, penumbral filaments (magneto)convection cells, as suggested by recent numerical modellings? 
	
High spatial resolution, precise, and high signal-to-noise observations by the Solar Optical Telescope \citep{2008SoPh..249..233I,2008SoPh..249..221S,2008SoPh..249..197S,2008SoPh..249..167T} on-board Hinode \citep{2007SoPh..243....3K} have extraordinarily contributed to understanding of sunspot structure and dynamics, in the first 10 years, by providing new information about many sunspot features, including umbral dots, light bridges, penumbral filaments, moat regions, and disclosed their internal structures. Hinode helped addressing several of the above-mentioned questions, and opened new directions. See \citet{2003A&ARv..11..153S} for a detailed review on sunspot structure and for open questions thereon before the Hinode-era.

In this section (\ref{sec71}), we review some of the latest developments, achieved from the data of unprecedented high quality obtained by Hinode, in establishing (mostly photospheric) thermal, flow and magnetic properties of sunspot structures both at small and global scales. Note that although works on umbral dots, light bridges, moving magnetic features, umbral/penumbral jets, and formation/decay of sunspots are reviewed, a more extensive detail is given on the fin-structure of the sunspot penumbra, the most complicated magnetic structure on the surface of the Sun, to understanding of which the Hinode has contributed the most significantly. We also discuss some questions that have emerged as a result of these new observations, i.e., about sunspot structure, dynamics and their connection with the upper atmosphere, and point out the need of multi-height/multi-temperature observations at a higher spatial resolution and cadence that are needed to answer them and that are anticipated from future generation solar telescopes e.g., DKIST and the next Japan-led solar space mission. 

For past reviews on the structure of sunspots, please see
\citet{1981SSRv...28..387M},
\citet{1981SSRv...28..435S},
\citet{1985ARA&A..23..239M},
\citet{1991GApFD..62..249S},
\citet{1997ASPC..118..155S},
\citet{2003A&ARv..11..153S},
\authorcite{2004ARA&A..42..517T}
(\yearcite{2004ARA&A..42..517T}, \yearcite{2008sust.book.....T}),
\citet{2009SSRv..144..229S},
\citet{2009ASPC..415..339T},
\citet{2011LRSP....8....4B}, and
\citet{2011LRSP....8....3R}.  
In recent years MHD simulations have made significant progress in reproducing many aspects of small-scale structures of sunspots \citep{1996ApJ...457..933H,2000MNRAS.314..793H,2006ApJ...641L..73S,2007ApJ...669.1390H,2008ApJ...677L.149S,2009ApJ...691..640R,2011LRSP....8....3R,2012ApJ...750...62R}. In this review, we mainly focus on the observational results and, when suitable, mention relevant simulations.

\section{Umbral dots and light bridges}
\label{sec711}
Sunspot umbrae often contain light bridges (LBs) and umbral dots (UDs); both are enhanced bright structures inside dark umbrae, magnetoconvection being a proposed mechanism of heat transport in them \citep{2002AN....323..371W,2006ApJ...641L..73S,2007PASJ...59S.585K,2009ApJ...702.1048W,2014PASJ...66S...1W}. Using Hinode/SOT-SP data, \citet{2008ApJ...678L.157R} detected upflows of 800~m~s$^{-1}$, and a field weakening of some 500~G in UDs; see also
\citet{2009ApJ...694.1080S} and \citet{2015SoPh..290.1119F} for a comparison of central and peripheral UDs. \citet{2013A&A...554A..53R} further analyzed the same sunspot data using a more sophisticated inversion technique and detected systematic diffuse downflows surrounding UDs, consistent with the downflows seen by \citet{2010ApJ...713.1282O} in a few UDs of a pore. \citet{2013A&A...554A..53R} further found that upflowing mass flux in central part of UDs balances well with the downflowing mass flux in their surroundings. Evidence of dark lanes in UDs, as predicted by MHD simulations of \citet{2006ApJ...641L..73S}, was reported by \citet{2007ApJ...669L..57B} and \citet{2008ApJ...672..684R}. On the other hand, \citet{2012ApJ...752..109L} and \citet{2013A&A...554A..53R} could not detect it, thus questioning the magnetoconvective nature of UDs. Furthermore, MHD simulations suggest concentrated downflows at the UD boundary, not found in observations so far.

LBs, often apparent as a lane of UDs, separate sunspot umbrae into two or more parts of the same polarity magnetic field. They can be divided into `granular' photospheric substructures (e.g., \cite{1991ApJ...373..683L,2010ApJ...718L..78R,2014A&A...568A..60L}), `faint' LBs \citep{1991ApJ...373..683L,2009A&A...504..575S}, or `strong' LBs \citep{2008ApJ...672..684R,2012A&A...537A..19R}. Similar to UDs, magnetic fields in all types of LBs are more inclined from vertical as compared to their surroundings \citep{2006A&A...453.1079J,2007PASJ...59S.577K,2014A&A...568A..60L,2016A&A...596A..59F}, and similar to the convergence of spine field  over penumbral filaments (described later), umbral field converges above LBs. Supporting their convective nature, upflows in central part of LBs and surrounding strong downflows have been observed
\citep{1997ApJ...490..458R,2002A&A...383..275H,2009ApJ...704L..29L,2010ApJ...718L..78R,2014A&A...568A..60L}. Dark lanes have been detected in LBs using Hinode data by e.g.,
\citet{2007ApJ...669L..57B} and \citet{2014A&A...568A..60L},
thus supporting the magnetoconvective nature of LBs. \citet{2014A&A...568A..60L} found field-free regions in granular LBs with similarities to ``normal" quiet-Sun granules, thus suggesting that, unlike other umbral features (i.e., UDs and other types of LBs), granular LBs could be made by convection from deeper layers. In a recent work by using Hinode SOT-SP time series of a sunspot, \citet{2018ApJ...852L..16O} found a LB to have the strongest magnetic field over the sunspot.

Several small-scale jet-like events in connection with UDs and LBs have also been reported using Hinode data
(e.g., \cite{2009ApJ...696L..66S,2011ApJ...738...83S,2014A&A...567A..96L,2015MNRAS.452L..16B,2015ApJ...811..137T,2016A&A...594A.101Y}).

\section{Structure of sunspot penumbral filaments}
\label{sec712}
With the presence of rapidly varying fine-scale field, flow, and thermal properties, both in radial and azimuthal directions, sunspot penumbra represents, undoubtedly, the most complicated and challenging structure on the solar surface. Penumbrae are made of copious thin bright filaments \citep{1993ApJ...403..780T,1995A&A...298..260R,2005A&A...436.1087L,2007Sci...318.1597I,2011LRSP....8....4B} and dark spines \citep{1993ApJ...418..928L}. See also
\citet{2009ApJ...697L.103S} and \citet{2009ApJ...702L.133T}
for fine-scale distribution of local magnetic twists and current densities in sunspot penumbrae, and
\citet{2009ApJ...700..199T} and \citet{2010ApJ...720.1281G}
for the effect of polarimetric noise in estimating these parameters using Hinode data.

According to theoretical expectations \citep{1953sun..book..532C,1958IAUS....6..263S,1977SoPh...55....3S,1994A&A...290..295J,2008sust.book.....T,2014masu.book.....P}, the presence of strong magnetic field of 1--2~kG should prohibit convection in sunspot penumbrae, thus, keeping them dark, similar to umbrae. As penumbrae have some 75\% brightness of quiet-Sun intensity, some form of convection takes place therein. It may be radial, i.e., upflows take place in the inner penumbrae and downflows in the outer penumbrae. Or there could be azimuthal/lateral convection, in that upflows take place all along the filament's central axis and downflows along the sides of the filament. Or the convection in penumbrae may be a combination of the above two \citep{2011LRSP....8....4B}. The presence of radial convection was evidenced by e.g.,
\citet{2006ApJ...646..593R},
\citet{2007Sci...318.1597I}, \citet{2013A&A...557A..25T}, and
\authorcite{2009A&A...508.1453F}
(\yearcite{2009A&A...508.1453F}, \yearcite{2013A&A...550A..97F}).
Support for azimuthal convection was found by
\citet{2007Sci...318.1597I},
\citet{2008A&A...488L..17Z},
\citet{2010ApJ...722L.194B},
\citet{2011ApJ...734L..18J},
\citet{2011Sci...333..316S}, \citet{2012A&A...540A..19S}, 
\citet{2013A&A...557A..25T}, and \citet{2015ApJ...803...93E},
while other researchers could not detect such downflows \citep{2009A&A...508.1453F,2010ApJ...725...11B,2010ApJ...720.1417P}. Furthermore, convection in the penumbra can take place in the presence of strong magnetic field \citep{2009ApJ...691..640R,2011LRSP....8....3R,2012ApJ...750...62R}, or in a very weak field or in absence of it (field-free gaps) [\cite{2006A&A...460..605S,2006A&A...447..343S}].

By using Hinode/SOT-SP data of a sunspot (leading-polarity sunspot of NOAA AR~10933) observed almost on the solar disk center ($\mu=0.99$) on 2007 January 5 (during 12:36--13:10~UT), \citet{2013A&A...557A..25T} explored the fine structure of penumbral filaments. \citet{2015A&A...583A.119T} then studied global properties of the same sunspot in light of the fine structure of filaments and spines, and sorted-out the thermal, velocity, and magnetic structures of the whole sunspot. In the following we summarize some of the main results found in these papers, with appropriate discussion and additional topics included. Interestingly, different aspects of Hinode data of this particular sunspot has been studied by several researchers, which has resulted in many other publications
(e.g.,
\cite{2008ApJ...681.1677K,2009A&A...508.1453F,2009PhDT.........8T},
\yearcite{2012ApJ...744...65T};
\cite{2009ApJ...702L.133T, 2009ApJ...706L.114V}, \yearcite{2010A&A...516L...5V};
\cite{2010A&A...524A..20K,2011LRSP....8....4B,2011PhDT.......137F,2013A&A...554A..53R,2013A&A...557A..24V,2017A&A...599A..35J}).

\begin{figure*}[htp]
	\includegraphics[width=\textwidth]{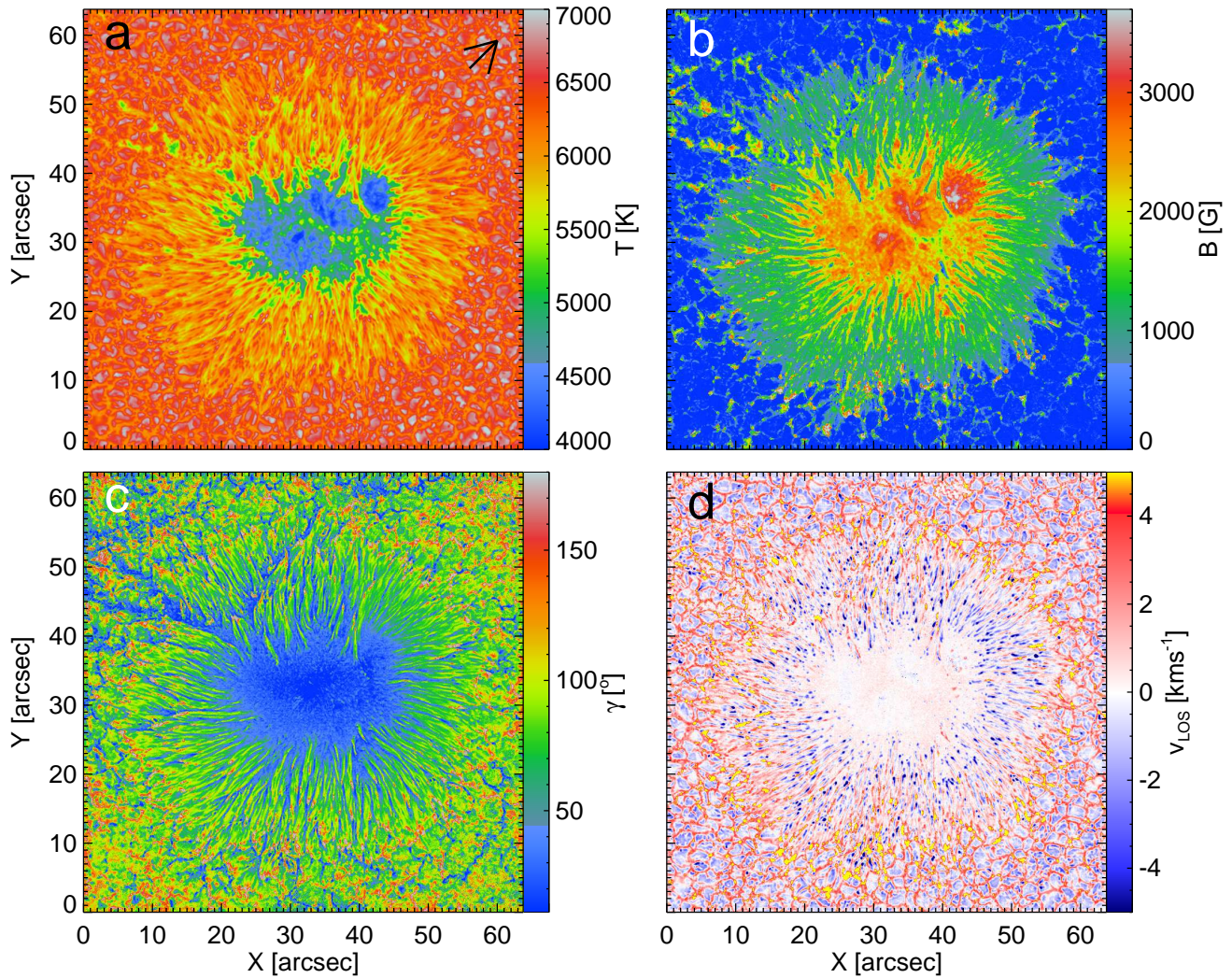}
	\caption{%
		Four selected maps of physical parameters of the leading positive magnetic polarity sunspot from AR~10933, observed by Hinode/SOT-SP and inverted using spatially coupled inversions. (a) $T$ map; a black arrow points to the solar disk center. (b) $B$ map. (c) $\gamma$ map. (d) $v_{\rm LOS}$ map. Color bars for the parameters are attached to the right of each panel of the figure, and are scaled to enhance the visibility of spatial variations in the parameters. [Reproduced from \citet{2015A&A...583A.119T} by permission of ESO.]
	}
	\label{7.1-fig2}
\end{figure*}

\paragraph*{\bf Inversion of Hinode (SOT/SP) data:} 
 Before we discuss new findings on small-scale structure of sunspots, we summarize here what sort of spectropolarimetric inversions were developed on Hinode SOT-SP data (due to known telescope parameters, e.g., \cite{2008A&A...484L..17D}), and used to get many of the results described later. To infer physical parameters from the observed Stokes profiles different Milne-Eddington (the simplest model) and depth-dependent inversions (based on response functions) have been traditionally used \citep{1987ApJ...322..473S,1992ApJ...398..375R,1998ApJ...507..470S,2000A&A...358.1109F,2004A&A...414.1109L,2008ApJ...683..542A}. Although we do not have to deal with the effects of the earths atmosphere in the spectropolarimetric data of Hinode, we do have to deal with the spatial and spectral degradation caused by the telescope and the detector. In particular the spectral degradation was taken into account by \citet{2007PASJ...59S.837O} considering the local straylight as a second atmospheric component, which was considered to be contributed by telescope diffraction and not by unresolved small-scale structure. In this method of inversions, a significant amount of the signal ($\sim$75\% due to telescope diffraction) gets subtracted from each pixel, thus, significantly reducing the signal-to-noise ratio of the results.  To properly take in to account the spatial degradation caused by the telescope diffraction, \citet{2012A&A...548A...5V} developed a new method, spatially coupled inversion, in which the spectropolarimetric data is degraded in a known way, using the telescope point spread function (PSF), and the atmospheric parameters over the whole field of view (FOV) of the data are simultaneously constrained.     

In the spatially coupled inversion, Stokes profiles for all pixels in a given FOV are synthesized and then convolved with the PSF of the telescope, and then these are matched with the observed Stokes profiles until $\chi$-squared is minimized; then physical parameters are inferred. This method allows accurate fitting of Stokes profiles over a large FOV, and improves signal-to-noise and spatial resolution of the inversion results. Further, the spatially coupled inversion can be carried out at a higher pixel resolution than that of the observed magnetogram by artificially refining the pixel grid of the solution, thus, resolving additional substructures down to the diffraction limit of the telescope, that were not resolved with earlier, pixel based, inversions of Hinode (SOT/SP) data. 
	
For exploring the internal structure of sunspot penumbra \citet{2013A&A...557A..25T} used this newly developed spatially coupled inversion code implemented in the SPINOR code \citep{2000A&A...358.1109F}, which returns depth-dependent physical parameters, based on their response functions to the used spectral lines. \authorcite{2013A&A...557A..25T} (\yearcite{2013A&A...557A..25T}, \yearcite{2015A&A...583A.119T}) used a pixel size of 0.08\arcsec and the structures down to the diffraction limit of the telescope were resolved. A three-node inversion was performed and the best results obtained after several experiments were used for analysis. The physical parameters returned from the inversion are temperature $T$, magnetic field strength $B$, field inclination $\gamma$, field azimuth $\phi$, line-of-sight velocity $v_{LOS}$, and a microturbulent velocity $v_{mic}$. Before the velocities were inferred, a velocity calibration was done by assuming that umbra, excluding UDs, is at rest. Maps of the sunspot in a few selected physical parameters from the inversion are shown in figure \ref{7.1-fig2}.

\paragraph*{\bf Selecting penumbral filaments:}
From the maps of the physical quantities returned from the inversions of the Hinode SOT-SP data of a sunspot, \citet{2013A&A...557A..25T} were able to isolate penumbral filaments. However, because a single parameter map was not sufficient to track full filaments, e.g., filament heads ( the ``head" of a filament is the part of the filament nearest to the sunspot umbra) were clearly visible in $T$ and $v_{\rm LOS}$ maps but could not be detected in $\gamma$ maps, and the tails (the ``tail" of a penumbral filament is the part of the filament farthest from the sunspot umbra) of filaments could not be detected in $T$ maps, \citet{2013A&A...557A..25T} combined $T$, $v_{\rm LOS}$, and $\gamma$ maps for selecting filaments. The selected penumbral filaments were de-stretched and straightened using bi-cubic spline interpolation and normalized to a certain length. To reduce fluctuations and to extract common properties to all filaments, they averaged filaments after sorting them into inner, middle, and outer filaments. Before the work of \citet{2013A&A...557A..25T}, the full picture of a penumbral filament was not known (see, e.g., \cite{2011LRSP....8....4B}.)

\begin{figure}[t]
	\includegraphics[width=1.01\columnwidth]{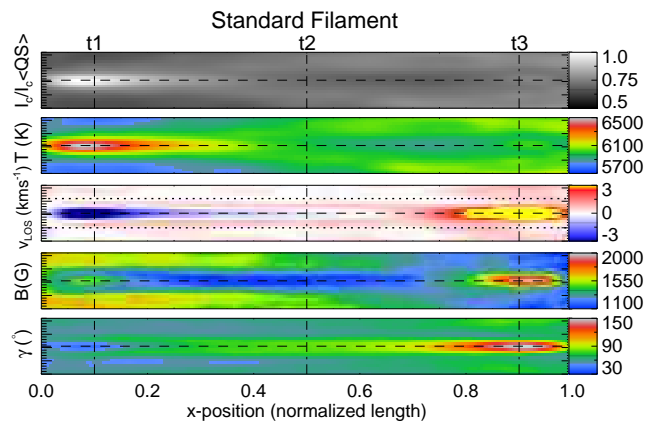}
	\caption{%
		Maps of five physical parameters of the standard penumbral filament, all at the surface of unit optical depth. Quantitative values along the longitudinal dashed line at the central axis of filaments and at three transverse cuts can be found in \citet{2013A&A...557A..25T}. The total width including the surrounding mostly spine field is $1\farcs 6$; the width of the filament itself, outlined for reference in the $v_{\rm LOS}$ map by two longitudinal dotted lines, is $0\farcs 8$. [Reproduced from \citet{2013A&A...557A..25T} by permission of ESO.]
	}
	\label{7.1-fig3}
\end{figure}

\paragraph*{\bf Uniformity of properties in all penumbral filaments, and the ``standard filament":}
The selected penumbral filaments showed similar spatial properties everywhere, in the inner, middle, and outer parts of the sunspot penumbra. Therefore, \citet{2013A&A...557A..25T} averaged all selected penumbral filaments to create a ``standard penumbral filament". In figure \ref{7.1-fig3}, we display a few physical parameters of the standard filament at the optical depth unity. Please see \citet{2013A&A...557A..25T} for the plots of their depth-dependence and quantitative properties.

\paragraph*{\bf Size of filaments:}
The lengths of filaments varied from $2''$ to $9''$ with an average of $5'' \pm 1\farcs 6$, whereas the width of each filament remained close to the averaged width of $0\farcs 8$.

\paragraph*{\bf Thermal properties, heads of filaments --- penumbral grains:}
All penumbral filaments contained a bright head in $I_{\rm c}$ and $T$ ($\sim$6500 K) maps at the optical depth unity, with a rapid fall in temperature (and intensity) along their central axes towards the tail, the difference in the temperatures of the heads and the tails reaching up to 800 K. The teardrop-shaped heads of penumbral filaments were earlier referred to as penumbral grains \citep{1973SoPh...29...55M,1999A&A...348..621S,2006ApJ...646..593R,2013A&A...560A..77Z}.

\paragraph*{\bf Dark lanes:}
A dark core along the central axis of the ``standard filament" is clearly visible in the middle and higher photospheric layers \citep{2013A&A...557A..25T};
see \citet{2002Natur.420..151S},
\citet{2007ApJ...668L..91B},
\citet{2007A&A...464..763L}, and
\citet{2008ApJ...672..684R}
for earlier reports of dark lanes in penumbral filaments. These can be as narrow as 0.1" \citep{2016A&A...596A...7S}.
The dark lanes are the locations of weak and more horizontal magnetic field than their surroundings, consistent with the observations of
\citet{2007ApJ...668L..91B} and \citet{2007A&A...464..763L}.
The weak field at these locations results in a higher gas pressure thus raising the optical depth unity surface to the higher and cooler layers, which are then visible as dark lanes \citep{2006A&A...447..343S,2007A&A...471..967B,2008A&A...488..749R}.

\paragraph*{\bf Magnetic field in penumbral filaments and convergence of surrounding spine field:}
With horizontal distance along a filament from its head, the field inclination changes from more vertically up ($\gamma\sim$10-40$^{\circ}$) in the head (where the field is strong), to horizontal in the middle (where the field is weaker), and then to downward ($\gamma\sim$140-170$^{\circ}$) in the tail (where the field is stronger), thus making an inverse U shape. What happens to the field when it dips down into the photosphere at the tails of filaments is not known. They could form a sea-serpent, bipolar structure \citep{2008A&A...481L..21S,2010AN....331..563S}, could remain below and disperse \citep{2013A&A...557A..25T}, or return back to the surface well outside the sunspot \citep{2002Natur.420..390T}.

The presence of more horizontal field in the middle of filaments at higher layers found by \citet{2013A&A...557A..25T} agrees with the inverse-U shape of penumbral filaments. The surrounding spine fields were found to diverge in the deepest layers and to converge together above the filament making a cusp shape, in agreement with the results of \citet{2008A&A...481L..13B}, who also analyzed Hinode data of a sunspot penumbra. The convergence of spine field with height over a filament agrees with the model of \citet{1993A&A...275..283S}.

\paragraph*{\bf Absence of evidence of field-free gaps in penumbral filaments:}
Magnetic field strength is weaker along the middle of a filament but still has a value of $\sim$1000~G \citep{2013A&A...557A..25T}. This indicates that the flow in filaments is not field-free, thus supporting the view that the Evershed flow is magnetized \citep{1994A&A...283..221S,2005A&A...436..333B,2008A&A...481L...9I,2012ApJ...750...62R}. In agreement with this result and with that of \citet{2008ApJ...687..668B}, recent deep-photospheric observations of sunspots in Fe {\sc i} lines, at around 1565~nm found no evidence of the regions with weak (B$<$500~G) magnetic fields in the sunspot penumbrae \citep{2016A&A...596A...2B}.

\paragraph*{\bf Convective nature of penumbral filaments:}
All penumbral filaments display a clear pattern of convection both in the radial and azimuthal directions; upflows concentrate in the head (at $\sim$5~{\kms}, on average) but continue along the central axis up to more than half of a filament. Strong downflows concentrate in the tail (at $\sim$7~{\kms}, on average) of each filament.  In addition weak but clear downflows (of 0.5 \kms) are visible along the side edges of penumbral filaments, see also \citet{2011ApJ...734L..18J,2011Sci...333..316S,2012A&A...540A..19S,2013A&A...549L...4R,2013A&A...553A..63S}, and \citet{2015ApJ...803...93E}. A scatter plot made by \citet{2013A&A...557A..25T} between $T$ and $v_{\rm LOS}$ revealed that upflows are systematically hotter than downflows by some 800~K, thus quantitatively supporting convective nature of penumbral filaments.

\paragraph*{\bf Opposite polarity magnetic field at the sites of lateral downflows:}
In 20 of the 60 penumbral filaments studied by \citet{2013A&A...557A..25T} the narrow downflowing lanes at the sides of filaments were found to carry opposite polarity magnetic field to that of spines and to that in the heads of filaments. Similar opposite polarity fields inside sunspot penumbrae were also reported by e.g.,
\citet{2013A&A...549L...4R},
\citet{2013A&A...553A..63S}, and
\citet{2016A&A...596A...4F}.
The opposite magnetic polarity field along the filament sides was averaged out in the standard penumbral filament in figure \ref{7.1-fig3}.

\paragraph*{\bf The Evershed flow:}
Consistent with the presence of dominant upflows in the inner penumbrae and dominant downflows in the outer penumbrae
(\cite{2009A&A...508.1453F};
\cite{2013A&A...557A..25T}, \yearcite{2015A&A...583A.119T}, \cite{2013A&A...557A..24V})
the Evershed flow can be explained as a siphon flow in magnetized horizontal flux tubes \citep{1968MitAG..25..194M,1993A&A...275..283S,1997Natur.390..485M,1998ApJ...493L.121S,2007PASJ...59S.593I,2014A&A...564A..91J}. However, the siphon flow was ruled out in the recent past due to the presence of stronger magnetic field in the inner penumbrae than the outer penumbrae, which is instead more suitable to drive an inverse Evershed flow (inflow, due to higher gas pressure in the outer penumbrae and beyond), and also due to the support to the alternative idea of convection driving naturally the Evershed flow guided by inclined magnetic field \citep{1996ApJ...457..933H,2008ApJ...677L.149S,2010ASSP...19..186I}.

An enhanced magnetic field (1.5--2~kG, on average) was seen in the heads, and even stronger field (2--3.5~kG, on average) was found in the tails of penumbral filaments at log$(\tau)=0$ by \citet{2013A&A...557A..25T}. This observation is consistent with a siphon flow to drive the Evershed flow, see also \citet{2017A&A...607A..36S}. However, because the geometrical heights of different parts of penumbral filaments are not known, no definite conclusion can yet be made. On the other hand the clear observation of both the radial and azimuthal convection supports the idea of
\citet{1996ApJ...457..933H} and \citet{2008ApJ...677L.149S}
that the presence of inclined field guides the convecting gas to generate an outflow, the Evershed flow. Moreover, the upflows being systematically hotter than the downflows in penumbral filaments support the idea that the gas rises hot near the head and along the central axis of a filament for more than half of its length, and is then carried outward along the horizontal magnetic field (as the Evershed flow) and across it in the azimuthal direction \citep{2013A&A...557A..25T}. The gas cools along the way before it sinks down at the side edges and in the tail of the filament. The Evershed flow does not stop abruptly at the outer boundary of sunspots but continues outside it in the moat region \citep{1994A&A...283..221S,2006A&A...454..975R,2008ApJ...680.1467S,2009ApJ...701L..79M}. 

\paragraph*{\bf Penumbral jets and bright dots:}
Penumbral jets are narrow transient bright events (10--20\% brighter than the surrounding background), discovered by \citet{2007Sci...318.1594K} using the Ca {\sc ii} H-line filter on Hinode/SOT-FG\@. They have lifetimes of less than a minute, widths of less than 600~km, lengths of multiple thousand km, and speeds of more than 100~\kms. These jets stream along the spine field, which get more inclined to the vertical with increasing horizontal radius in the penumbrae \citep{2008A&A...488L..33J,2015A&A...583A.119T}. Some of these jets heat the transition region directly above, but quantifying their coronal contribution requires further investigation \citep{2016ApJ...816...92T}.

\begin{figure}
	\includegraphics[width=1.03\columnwidth]{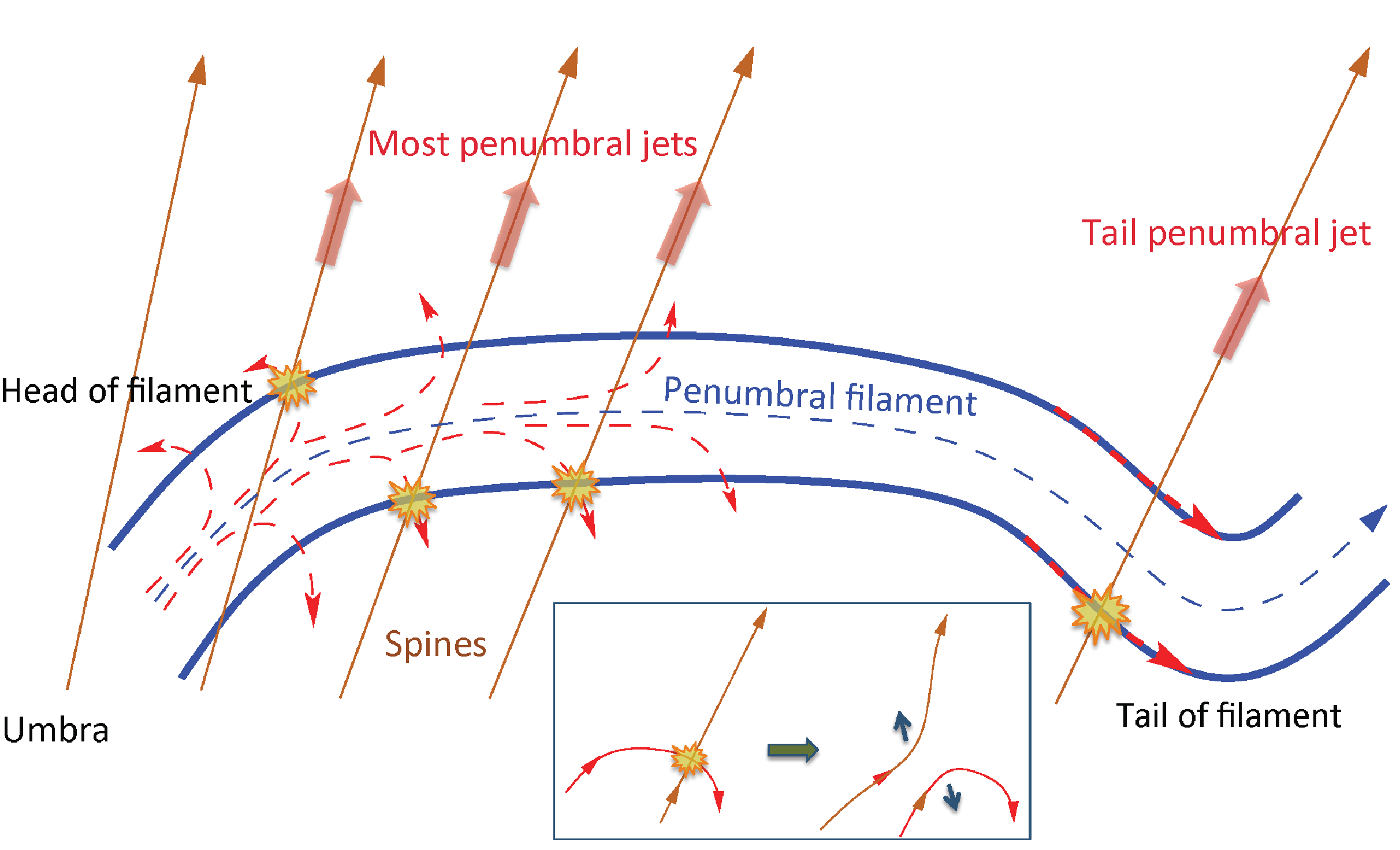}
	\caption{%
		A schematic sketch (not to scale) illustrating the formation of sunspot penumbral jets. All jets travel along the spine fields, which are more vertical in inner penumbrae (near filament heads). The red dashed lines with arrow heads show the direction of field lines in the filament. In a box in the middle bottom the magnetic configuration as well as the reconnection of spine field with opposite polarity field at the filament edge are shown. [Reproduced from \citet{2016ApJ...816...92T}.]
	}
	\label{7.1-fig4}
\end{figure}

Based on the new complete picture of penumbral filaments \citep{2013A&A...557A..25T}, \citet{2016ApJ...816...92T} proposed a modified view of formation of penumbral jets. The magnetic reconnection can take place between the spine field and the opposite polarity field in the sides of filaments, due to the obtuse angle between them, partly in agreement with the numerical modeling of \citet{2008ApJ...687L.127S} and \citet{2010ApJ...715L..40M}, rather than a component reconnection taking place between the spine fields of the same magnetic polarity and having an acute angle between them. A cartoon diagram of this possibility is shown in figure \ref{7.1-fig4}. Similar but more repetitive and larger jets at the tails of penumbral filaments were also detected by \citet{2016ApJ...816...92T} using Hinode/SOT-FG data.

Other dynamic events in sunspot penumbrae include moving bright dots, recently discovered by \citet{2014ApJ...790L..29T} using IRIS data. Penumbral bright dots were also seen in Hi-C data \citep{2016ApJ...822...35A}.
\index{Hi-C: High Resolution Coronal Imager} Some of the bright dots and penumbral jets could be linked with each other and might have the same origin \citep{2016ApJ...829..103D,2016ApJ...816...92T,2017ApJ...835L..19S}; however, this subject requires extensive further investigation.

\section{Long-lived controversies resolved}
\label{sec713}
By exploring the complete picture of penumbral filaments using Hinode/SOT-SP data, and discovering that the physical properties of filaments change along their length, many of the long-standing controversies about structure of sunspot penumbrae are resolved. For example, the brightness and temperature of the downflowing regions can easily be confused with spines; both are darker regions than the heads of filaments. \citet{1993ApJ...418..928L} found more vertical fields/spines to be darker whereas \citet{2001ApJ...547.1130W} and \citet{2005A&A...436.1087L} found the spines to be warmer. This could be because the heads of filaments were mistaken to be spines, both of these having similar field inclination. Similarly, by looking at different parts of filaments \citet{2011LRSP....8....4B} concluded that the inter-spines are brighter filaments in the inner penumbrae and darker filaments in the outer penumbrae. The controversy also extends to whether the Evershed flow mainly takes place in the brighter or the darker regions of the penumbra \citep{1990ApJ...355..329L,1993ApJ...403..780T,2001A&A...378.1078H,2001ApJ...547.1148W}. However, from the fact that the upflows near heads are brighter and the downflows near tails are darker, one can interpret that the gas cools down as it travels along the filament central axis; thus the Evershed flow might be a natural outflow along the arched field. See \citet{2003A&ARv..11..153S} for detailed literature on several such controversies and \citet{2013A&A...557A..25T} for their clarifications, thus highlighting the importance of resolving the complex magnetic, thermal, and flow structure of filaments for correctly interpreting observations of sunspot penumbrae.

\section{Global properties of sunspots}
\label{sec714}
Hinode data confirmed and clarified several global properties of sunspots found in the past and added new information; e.g., in the past magnetic field canopy was found by different authors to start at different locations in penumbrae (e.g., \cite{2011LRSP....8....4B}). It was verified by \citet{2015A&A...583A.119T} that the canopy starts only at the outer visible boundary of sunspots, in agreement with the results of
\citet{1980SoPh...68...49G},
\authorcite{1992A&A...263..339S}
(\yearcite{1992A&A...263..339S},
\yearcite{1999A&A...347L..27S}), and
\citet{1993SoPh..148..201A}.

\paragraph*{\bf Penumbral spines and filaments:}
Spines have denser, stronger, and more vertical magnetic field in the inner penumbra. The spine field becomes less dense, less strong, but more inclined radially outward from the umbra. A comparison of scatter plots between $B$ and $\gamma$ for full sunspot and for only penumbral pixels revealed that spines have the same magnetic properties (except that these are more inclined) as the field in umbrae. Thus, \citet{2015A&A...583A.119T} concluded that spines are intrusions of umbral field into penumbrae. These locations of spines were consistently found to be locations of more force-free photospheric magnetic fields than elsewhere in sunspot penumbrae \citep{2012ApJ...744...65T}.

Further, a qualitative similarity between scatter plots of different parameters for the standard penumbral filament (including its surrounding spines) and for sunspot penumbra led \citet{2015A&A...583A.119T} to conclude that sunspot penumbra is formed entirely of spines and filaments; no third component is present.

\paragraph*{\bf Peripheral strong downflows:}
Hinode observations showed the presence of systematic strong, often supersonic, downflows at the outer penumbral boundary of sunspots, with the presence of opposite polarity field therein to that of umbra and spines (e.g., \cite{2007PASJ...59S.593I,2009A&A...508.1453F,2009ApJ...701L..79M,2013A&A...557A..24V,2015A&A...583A.119T}), but also see
\citet{2010A&A...524A..21J} and \citet{2010A&A...524A..20K}
for a different kind of flow reported in sunspot penumbrae. The strong peripheral downflows could be considered as the continuation of the Evershed flow outside sunspots \citep{1994A&A...283..221S,2009ApJ...701L..79M}.

\citet{2013A&A...557A..24V} discovered the presence of the strongest magnetic fields and LOS velocities, ever reported in the photosphere, exceeding 7~kG and 20~\kms, respectively, in a few locations at the periphery of sunspots. They found a linear correlation between the downflow velocities and the field strength, which was in good agreement with MHD simulations. Possibly these peculiar downflows are induced by the accumulation and intensification of penumbral magnetic field by the Evershed flow. This is implied by the finding that these locations of strong downflows at the periphery of sunspots were the locations where tails of several penumbral filaments converge \citep{2013A&A...557A..25T, 2013A&A...557A..24V}.

\paragraph*{\bf Field gradients in sunspots:}
Generally, the field strength in sunspots decreases with increasing horizontal radius and height \citep{2001ApJ...547.1130W,2003A&A...410..695M,2011LRSP....8....4B,2015A&A...583A.119T}. A decrease in the average field strength from 2800~G in umbra to 700~G at outer penumbral boundary in the deepest layers was found in a sunspot observed by Hinode \citep{2015A&A...583A.119T}. The sunspot umbra showed an average vertical field gradient of 1400~G km$^{-1}$ in the deepest layers, which drops rapidly with height, reaching to 0.95~G km$^{-1}$ at log($\tau$) = $-2.5$.

However, in addition to the canopy structure seen at the outer penumbral boundary, an inverse field gradient (field increasing with height) was found in the inner-middle penumbrae \citep{2015A&A...583A.119T}. \citet{2017A&A...599A..35J} investigated this particular property of sunspots in details. They also found the presence of inverse gradient in MHD simulations. A closer look revealed the dominance of inverse gradients near the heads of penumbral filaments. The observed inverse field gradient could be a result of spine fields converging above filaments, the Stokes $V$ signal cancellation at filament edges, or an artefact caused by highly corrugated optical depth unity surface in the inner penumbrae \citep{2015A&A...583A.119T,2017A&A...599A..35J}. See a recent review on the height dependence of magnetic fields in sunspots by \citet{2018SoPh..293..120B}.

\begin{figure}
\includegraphics[width=1.01\columnwidth]{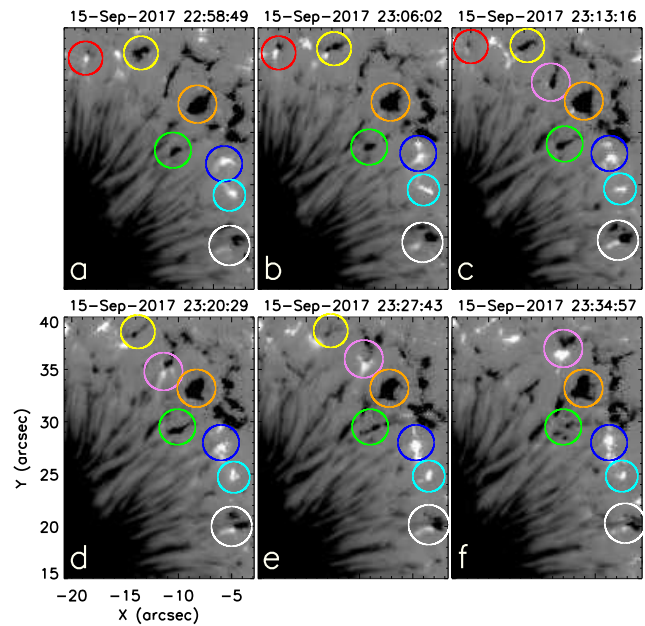}
\caption{%
	Line-of-sight magnetic field maps of the upper right quarter of a sunspot penumbra, including its surrounding moat region, observed by SOT/SP in normal scan mode, thus having a pixel size of 0.16". Evolution of eight moving magnetic features are outlined by circles, each in a different color: red (panels a--c), yellow (panels a--e), orange (panels a--f), violet (panels c--f), green (panels a--f), blue (panels a--c), cyan (panels a--f), and ivory (panels a--f). The SP data used in this figure were inverted at the Community Spectropolarimetric Analysis Center ({http://www2.hao.ucar.edu/csac}). 
}
\label{7.1-fig5}
\end{figure}

 \paragraph*{\bf Moving magnetic features and sunspot decay:}
Moving magnetic features (MMFs, see figure \ref{7.1-fig5}) are small, unipolar or bipolar, structures of sizes $< 2$\arcsec\ and lifetimes of 10 minutes to 10 hours. These move radially outward starting from the sunspot penumbra (or from within the moat region) with speeds of $< 2$ \kms\ and eventually disappear in the network fields \citep{1969SoPh....9..347S,1973SoPh...28...61H,1988SoPh..115...43B,2005ApJ...635..659H,2006SoPh..237..297R,2008A&A...481L..21S,2012ApJ...753...89L,2013ApJ...771...22L}. Although MMFs are prominent sunspot features, they are also found in pores \citep{2009A&A...500L...5Z,2012A&A...546A..26C,2012A&A...538A.109V,2017ApJS..229...13K}. Using Hinode SOT magnetograms \citet{2013ApJ...771...22L} found that half of MMFs in a sunspot are produced withing the penumbra and the other half originate within the moat region. They found that most of the MMFs formed in the moat are due to flux emergence. Once MMFs are formed, they start decaying by flux cancellation. The Evershed flow has been linked with the formation of MMFs \citep{2002AN....323..342M,2007A&A...471.1035Z,2008ApJ...686.1447K,2015ApJ...814..125R}, but for disagreements, see \citet{2013A&A...551A.105L}.

MMFs are proposed to play a crucial role in the decay of sunspots \citep{1973SoPh...28...61H,2002AN....323..342M,2005ApJ...635..659H}. Consistent with the results of \citet{2005ApJ...635..659H}, using Hinode SOT data \citet{2008ApJ...681.1677K} and \citet{2008ApJ...686.1447K} showed that in decaying sunspots the rate of the loss of magnetic flux (8$\times10^{15} Mx ~s^{-1}$) in sunspots is very similar to the rate of the magnetic flux carried out by MMFs outwards, thus taking several weeks for a sunspot of 10$^{22}$ Mx to completely decay. \citet{2008ApJ...686.1447K} also showed that positive and negative polarities balance each other in the moat region, suggesting that most of the sunspot flux is transported to the moat region and then outward by MMFs, and then removed by flux cancellation in the network regions. The rate of flux transport by moat flows is consistent with that found in recent MHD simulations of \citet{2014ApJ...785...90R} and \citet{2015ApJ...814..125R}.

\paragraph*{\bf Sunspot formation: }
The formation of sunspots, being a subsurface process \citep{1955ApJ...121..491P}, remains observationally more poorly understood than its decay. Sunspots form as a result of coalescence of small emerging magnetic elements \citep{1985SoPh..100..397Z}. Consistently, in the MHD simulations of \citet{2010ApJ...720..233C,2012ApJ...753L..13S} and \citet{2014ApJ...785...90R} flux emergence in the form of fragmented flux tubes (caused by subsurface convection) coalesce by horizontal inflow to make sunspots.  

Much observational work has been devoted to penumbra formation. After a critical magnetic flux for umbra is reached any new flux joining the spot probably contributes to the formation of penumbra \citep{2010A&A...512L...1S}. Using Hinode data \citet{2012ApJ...747L..18S} found that an annular feature in Ca II H in the form of a magnetic canopy surrounding the umbra in the chromosphere plays a role in the formation of penumbrae, thus proposing that the knowledge of chromospheric magnetic field is essential to understand the formation mechanism of sunspot penumbra. \citet{2014PASJ...66S..11K} concluded, again by using Hinode data, that penumbra can form in a few different ways, e.g., by active accumulation of magnetic flux, or by a rapid emergence of new magnetic flux, or by appearance of twisted or rotating magnetic tubes. The formation of sunspot penumbra is still not fully understood, and apparently depends on various factors, e.g., field strength, field inclination, size, amount of flux  \citep{1998ApJ...507..454L,2010A&A...512A...4R,2012A&A...537A..19R,2014PASJ...66S..11K,2015A&A...580L...1J,2016ApJ...825...75M,2017A&A...597A..60J,2017ApJ...834...76M}.

\section{Summary and Future prospects}
\label{sec715}

Sunspot physics has seen a major revolution in the first decade of the Hinode-era.  Unprecedented observations of sunspots by the Hinode SOT have revealed or clarified several small-scale aspects of sunspots, especially umbral dots and light bridges in the umbra, filaments, spines and jets in the penumbra, field gradient inversions in the inner penumbra, MMFs and peripheral downflows in the outer penumbra.  
	
Hinode has solved several of the open questions that existed before the Hinode-era. Some of the most striking discoveries are umbral dots having dark lanes, and magnetoconvective flows in UDs with the balanced mass-flux showing striking similarities with MHD models, granular light bridges having field free regions, internal structure of penumbral filaments, spines and filaments being the only components in the penumbra, MMFs being compatible with the idea of them being responsible for sunspot decay. The most striking new results are for the sunspot penumbra. Penumbral filaments are found to be elongated magnetized convective cells \citep{2013A&A...557A..25T}, qualitatively supporting recent magnetohydrodynamic (MHD) simulations \citep{2012ApJ...750...62R}. Several small-scale features were found to be part of penumbral filaments, e.g., penumbral grains are found to be the heads of filaments. Penumbral spines are observed to be true outward extension of umbral field. Sunspot penumbrae are formed entirely of spines and filaments \citep{2015A&A...583A.119T}. 

Some enduring controversies about the complex penumbral structure, e.g., whether strands of more vertical field (spines) are warmer or cooler than strands of more horizontal field, whether the Evershed flow mainly takes place in dark or bright penumbral strands or there is no correlation between flow and brightness, whether more horizontal fields are found in darker or brighter penumbral regions, etc. (see for details \cite{2003A&ARv..11..153S}), have been resolved by uncovering the fact that spines and parts of filaments have some properties in common \citep{2013A&A...557A..25T}. 
A few of the unexpected discoveries about sunspots, using Hinode SOT data, include the magnetic field at the tails of penumbral filaments being stronger than that in the heads of penumbral filaments by 1--2 kG \citep{2013A&A...557A..25T}, the strongest magnetic field in many sunspots being found not in dark sunspot umbra, but rather often in light bridges \citep{2018ApJ...852L..16O}, or at the periphery of sunspots \citep{2013A&A...557A..24V}.

Now we briefly mention some of the problems that should be addressed in future, i.e., by using future generation telescopes e.g., DKIST and SOLAR-C.

Concentrated downflows (with opposite polarity magnetic field in them to that of umbral field) surrounding umbral dots are expected from MHD simulations but have not been detected so far, probably because of insufficient spatial resolution of currently available magnetic field data. Absence of such concentrated downflows and opposite polarity field in higher resolution data would challenge present MHD simulations.

Because of the limited temporal cadence of spectropolarimetric data with Hinode, the lifetime of several small-scale features (e.g., penumbral filaments) remains poorly estimated. Further, how penumbral filaments form, evolve, and interact with spines remains to be explored. Filaments and spines could result from loading/unloading of convecting gas onto/from the spine field. Or spines in penumbra could be a result of the overturning convection taking place in-between them. Once the vertical magnetic field is sufficiently weak and the field is sufficiently inclined, a sub-surface convective instability within the sunspot can perhaps take place to form a penumbral filament. To address the above, we need to follow a sunspot penumbra of decent size in higher temporal and spatial resolution spectropolarimetric data for a couple of hours or more. Probably the formation mechanism of filaments and their interaction with spines also hold the answer to the formation mechanism of PJs and BDs, which may contribute to coronal heating above sunspots (\cite{2016ApJ...816...92T,2016ApJ...822...35A}, and references therein).

Multi-height spectropolarimetric data are needed to provide a 3-D picture of sunspots. A recent study by \citet{2016A&A...596A...8J} showed the presence of fine-scale magnetic structure in the azimuthal direction in the upper chromospheric layers of sunspot penumbrae, consistent with that found in the photosphere, albeit with reduced amplitudes. Moreover, to understand the force balance in sunspots and their equilibrium (e.g., \cite{2010A&A...516L...5V,2010ApJ...720.1417P,2012ApJ...744...65T}), we need to develop a technique to estimate accurately the geometrical heights of different small-scale features in sunspots.

\begin{ack}
SKT would like to thank Dr. Ron Moore for helpful comments. Hinode is a Japanese mission developed and launched by ISAS/JAXA, with NAOJ as a domestic partner and NASA and STFC (UK) as international partners. It is operated by these agencies in cooperation with ESA and NSC (Norway). SKT's research was supported by an appointment to the NASA Postdoctoral Program at the NASA Marshall Space Flight Center, administered by Universities Space Research Association under contract with NASA. S.K.T. gratefully acknowledges his current support by NASA contracts NNG09FA40C (IRIS), and NNM07AA01C (Hinode).
\end{ack}

\label{ref}

\end{document}